\documentclass[twoside]{JINST}
\usepackage{graphicx}
\usepackage{rangecite}
\usepackage{amsmath}
\usepackage[figuresright]{rotating}
\usepackage{afterpage}

\title{30 kV coaxial vacuum-tight feedthrough for operation at cryogenic temperatures}

 \author{I.~Kreslo\thanks{Corresponding author.}, I.~Badhrees, S.~Delaquis, A.~Ereditato, S.~Janos, M.~Messina, U.~Moser, B.~Rossi, M.~Zeller \\ \llap{}Albert Einstein Center for Fundamental Physics, \\
 Laboratory for High Energy Physics, \\
 University of Bern, \\
 Sidlerstrasse~5,\\
 3012 Bern, Switzerland \\
}
\abstract{In this paper we describe the technology of building a vacuum-tight high voltage feedthrough which is able to
operate at voltages up to 30~kV. The feedthrough has a coaxial structure with a grounded sheath which makes it capable
to lead high voltage potentials into cryogenic liquids, without risk of surface discharges in the gas phase above the liquid level.
The feedthrough is designed to be used in ionization detectors, based on liquefied noble gases, such as Argon or Xenon.}

\keywords{HV feedthrough, Cryogenic detectors;  Liquid Argon; TPC}

\begin{document}
\section{Introduction}

Time Projection Chambers (TPC) based on liquefied noble gases utilize the drift of the ionization charge, produced by charged particles, 
to obtain information about the third coordinate of the particle track, thus allowing 3-D reconstruction of the event.
Typical drift fields in such chambers vary from a few hundreds V/cm to more than 40~kV/cm in specific cases. In order to 
create such electric fields a high voltage potential must be conducted  from the power supply through the cryogenic dewar flange into the liquid phase to the chamber electrodes. The flange is typically kept at the room temperature, while the detector medium may have
temperatures down to 77~K. The dewar containing the medium must be vacuum-tight in order to keep the medium pure.
Between the warm flange and the liquid surface there is a layer of gas. The feedthrough is required to be shielded from outside in such a way
that high electric field does not penetrate into the gas phase. Contrary would result in  dark or glow discharges at the surface of the feedthrough
leading in turn to an untolerable level of electro-magnetic interference (EMI) noise on the detector readout system. 
%The length of the feedthrough is usually about 50~cm from below or more because of thermal conduction considerations.

The typical design, used for instance by ICARUS detector \cite{Icarus} comprises a coaxial structure, with a rigid metal central conductor of a few millimeters in diameter,
a grounded metal tube of a few centimeters in diameter and an insulator in between them. Since during its operation the device is exposed to extremely high
temperature gradients, all components must have a very similar thermal expansion coefficient. Moreover, the dielectric must me able to
keep its integrity at temperatures down to 77~K without cracking. Cracking leads to an immediate electric breakdown, since cracks will
be filled with the noble gas very fast and the electric field inside the crack easily exceeds 10~kV/cm.
%Most of commonly used polymers have a thermal expansion coefficient in the range of $5-13\times10^5~K^{-1}$, while
%metals are in the range of  $2-3\times10^5~K^{-1}$. 
Precise matching of the thermal expansion factors of the central electrode, the ground tube and the insulating
polymer is a challenging task. 

We propose a different approach to this problem. Instead of matching thermal expansion factors we produce the feedthrough out of highly elastic
components,  allowing good mechanical stability even with moderately matched thermal properties.
The design presented in this paper is part of a comprehensive R\&D program on liquid Argon TPCs that we are conducting in Bern since a few years 
\cite{Ereditato1,Ereditato2,Ereditato3,nitroTPC,Biagio,Biagio2,LARN10}.

\section{Feedthrough design and production technology}
\begin{figure}[htbp]	
\center\includegraphics[angle=0, width=0.99\textwidth]{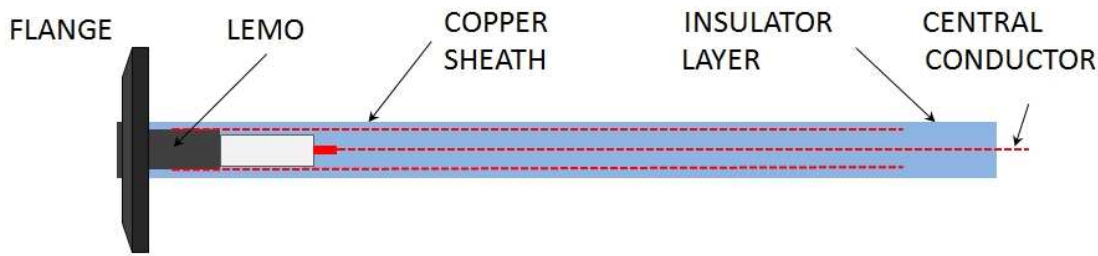}
\caption{Sketch of the feedthrough internal structure.}\label{struct}
\end{figure}

The inner structure of the proposed feedthrough is shown in Figure \ref{struct}. As an insulating polymer a double-component polyurethane resin\footnote{ARATHANE CW5620 BLAU by Huntsman Advanced Materials (Switzerland) GmbH,
Klybeckstrasse 200, CH-4057, Basel, Switzerland. } us used. This material is
rated to be mechanically stable and relatively elastic at temperatures down to 77~K. In order to reduce the tension between
the inner conductor and the insulating layer due to different thermal expansion at low temperatures the inner conductor
is made of bunch of twisted thin copper wires 0.3~mm diameter each. The outer diameter of the inner conductor
is about 2~mm, the twisting step is about 1~cm. This conductor has very low rigidity in longitudinal direction, and easily
absorbes the required longitudinal deformation together with the insulating layer.

The ground screen is made of a copper multi-wire sleeve with 30~mm in diameter, usually used to screen electric cables. The diameter of the wires the sleeve is made of is 0.2~mm.
Thus the sleeve is also easily deformable in both longitudinal and transversal direction. The sleeve not only covers the inner dielectric layer,
but is embedded into its surface so that no conductor is exposed to the outside. This approach helps to avoid any minor discharge to 
occur in the noble gas phase inside the detector.

\begin{figure}[htbp]	
\center\includegraphics[angle=0, width=0.3\textwidth]{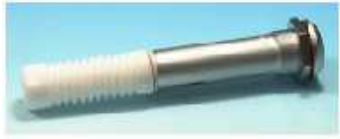}
\caption{High voltage LEMO connector rated for 30kV DC. The female part only is shown.}\label{lemo}
\end{figure}

The technology of the feedthrough manufacturing consists of the following steps. The 1~m long clean twisted central conductor is
soldered to a HV connector, rated for 30~kV operation\footnote{LEMO ERA.3Y.430.CTL} (see Figure \ref{lemo}). The assembly is then
fixed at the center of the 30~cm inner diameter 1~m long polyethylene tube with a spacer at the end opposite to the connector. The tube is pre-cut into two pieces along the axis
and reassembled with the aid of adhesive tape to ease the further disassembling. The tube with the central conductor and the connector is then 
put vertically and filled with liquid polyurethane resin, assuring that no air bubbles are captured inside. This is done with a syringe through a number of holes, 
made in the tube side surface. To provide reliable vacuum tightness, the open end of the conductor is slightly untwisted and twisted back to
ensure the good adhesion of the resin to the conductor wires and the absence of air channels.

Once the resin is cured the tube is disassembled into two pieces and the insulating stick with the central conductor and the connector is extracted.
As the next step the unit is placed horizontally at the stage of the lathe.  A layer of the liquid polyurethane resin is 
deposited by a brush onto the side of the insulating stick. An outer conductor sleeve is placed onto it and connected to
the ground surface of the HV connector. The sleeve is then fixed to the connector end with the wound metal wire, pulled and fixed in the same way
at the open end. 

\begin{figure}[htbp]	
\center\includegraphics[angle=0, width=0.99\textwidth]{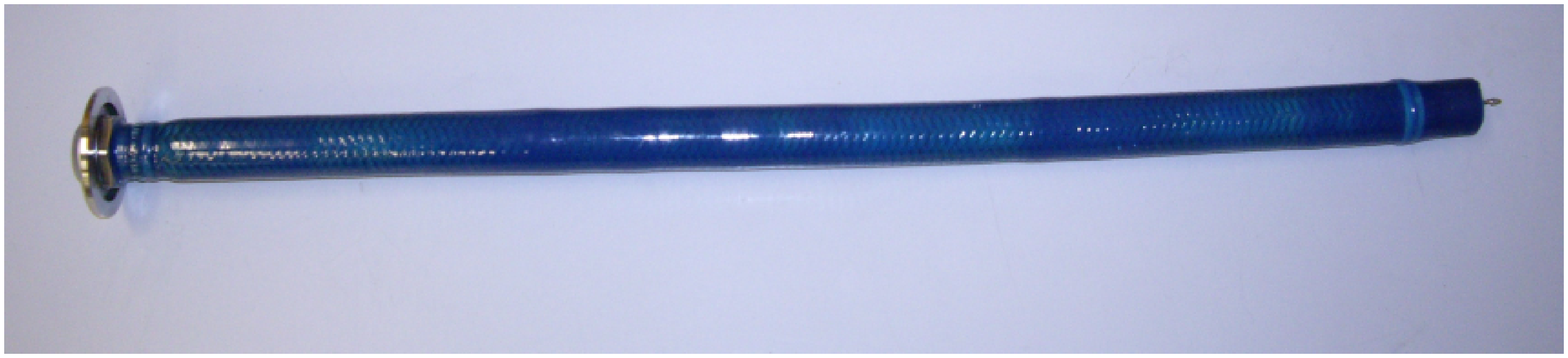}
\includegraphics[angle=0, width=0.49\textwidth]{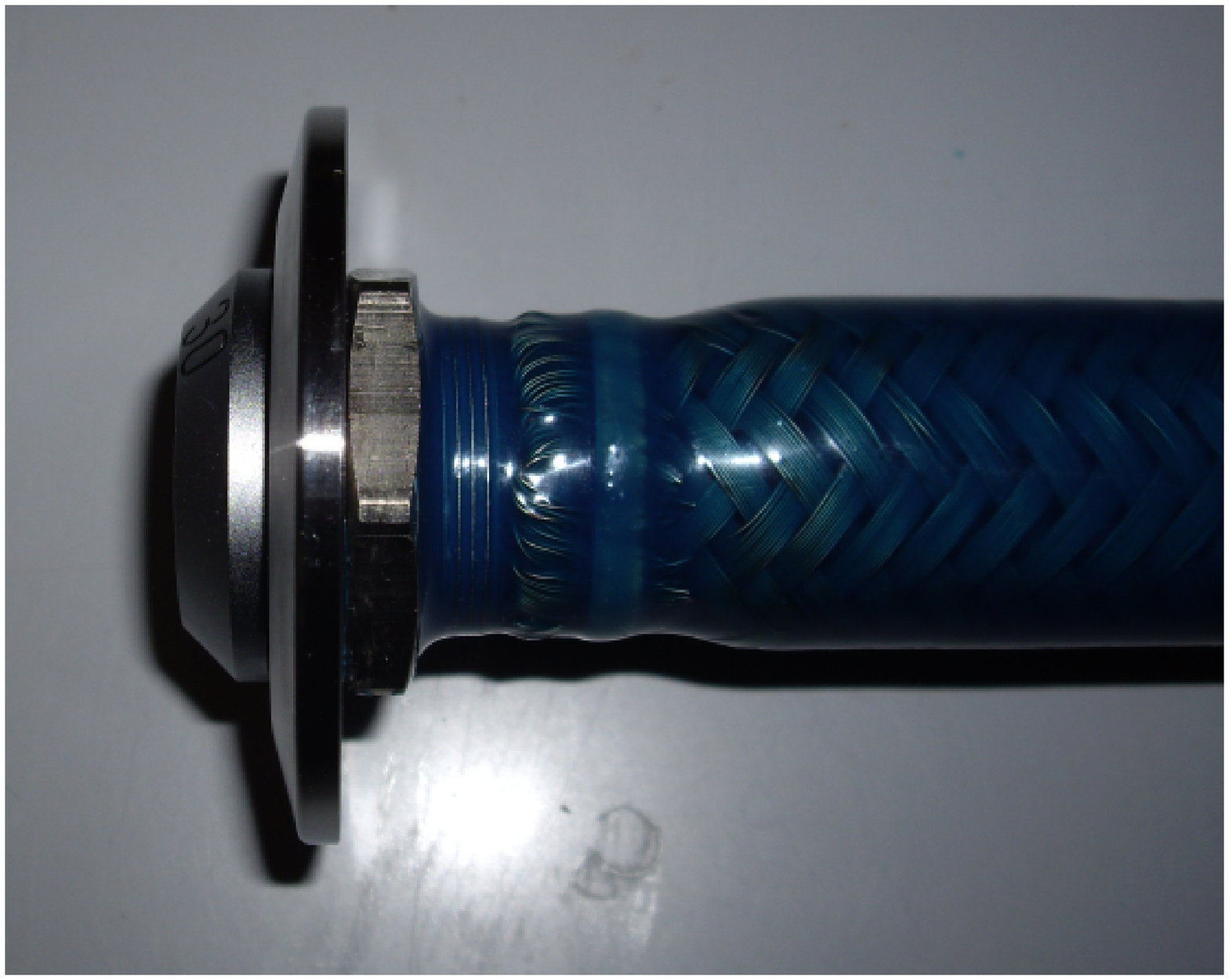}
\includegraphics[angle=0, width=0.50\textwidth]{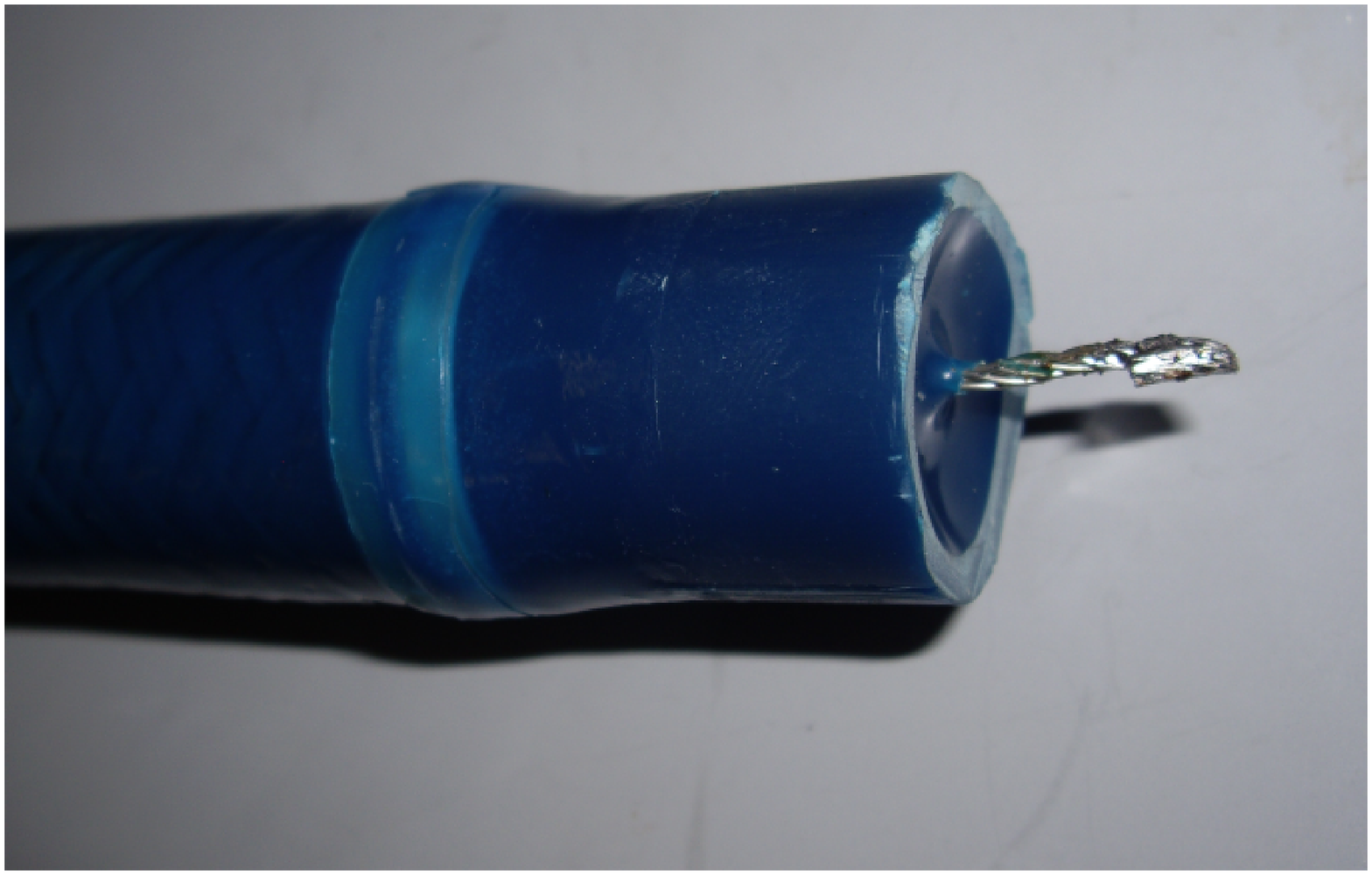}
\caption{Photo of the assembled feedthrough, 70 ~cm long.}\label{fig1}
\end{figure}

The unit is put into a
slow ($\approx$60~rpm) rotation and another layer of the liquid polyurethane resin is deposited by a brush onto the sleeve. Slow rotation 
provides even distribution of the resin on the unit surface. Once the resin is cured, the rotation is stopped and the feedthrough is 
released. The resulting unit and its details are shown in Figure \ref{fig1}. 
Total curing time of the given resin is about 48 hours. After this period the feedthrough is ready to be mounted in a detector.

\section{Conclusions}
A vacuum-tight coaxial high voltage feedthrough, capable to lead a voltage up to 30~kV into cryogenic liquids is designed and constructed.
The structure of the device allows to bring high potential into the liquid at low temperature without the risk of
surface discharges, which are very critical for low-noise ionization detectors, based on liquefied noble gases. The designed feedthrough
was successfully used in the 30~liter TPC with liquid Argon \cite{Biagio}, the 1~liter TPC with liquid Argon-Nitrogen mixture \cite{nitroTPC,LARN10,Marcel} 
and in a test setup for R\&D on 
new charge readout schemes for cryogenic ionization detectors \cite{Sebastien}. 
The unit was tested for vacuum tightness with a helium leak detector with the  sensitivity 
$10^{-10}$ mbar$\cdot$l/s and has shown no leak. 
%The unit was tested for vacuum tightness with a Helium leak tester and has shown no leaks at pressures down to $10^{-5}$~bar.
The feedthrough has demonstrated high stability under 
multiple cooling-heating cycles and at voltages up to 30~kV. We propose this design for the use in cryogenic TPCs based on liquefied
noble gases.

\section*{Acknowledgments}
We wish to thank all our technical collaborators for their significant contribution to the presented design, namely 
F.~Nydegger, J.~Christen. We would also express gratitude to our engineer staff: H.-U.~Sch\"utz and R.~H\"anni.

\end{document}